\documentclass[12pt,preprint]{aastex}

\def\be{\begin{equation}}
\def\ee{\end{equation}}
\def\bea{\begin{eqnarray}}
\def\eea{\end{eqnarray}}

\usepackage{graphicx}

\begin{document}

\title{SWIFT J164449.3+573451: a plunging event with the Poynting-flux dominated outflow}

\author{Wei-Hong Gao$^{1,2}$}
\affil{$^1$ Department of Physics and Institute of Theoretical
Physics, Nanjing Normal University, Nanjing, 210046, China.}
\affil{$^2$ Key Laboratory of Dark Matter and Space Astronomy,
Chinese Academy of Sciences, Nanjing, 210008, China.}
\email{gaoweihong@njnu.edu.cn}

\begin{abstract}
Swift J164449+573451 is a peculiar outburst which is most likely
powered by the tidal disruption of a star by a massive black hole.
Within the tidal disruption scenario, we show that the periastron
distance is considerably smaller than the disruption radius and the
outflow should be launched mainly via magnetic activities (e.g.,
Blandford-Znajek process) otherwise the observed long-lasting X-ray
afterglow emission satisfying the relation $L_{X}\propto\dot{M}$ can
not be reproduced, where $L_{X}$ is the X-ray luminosity and
$\dot{M}$ is the accretion rate. We also suggest that
$L_{X}\propto\dot{M}$ may hold in the quick decline phase of
Gamma-ray Bursts.
\end{abstract}


\setlength{\parindent}{.25in}

\section{INTRODUCTION}
Swift J164449.3+573451 (Sw J1644+57) triggered the Swift Burst Alert
Telescope (BAT) on 2011 March 28\citep{Cummings11, Burrows11}. As
revealed by the late optical observations, this transient lay at the
center of a galaxy at the redshift $z=0.3534$ \citep{Levan11a,
Levan11b}. In the first few weeks, the average isotropic luminosity
in the 0.3-10 keV band was about $10^{47}-10^{48} \rm erg s^{-1}$.
Several months later, it was still about a few $\times 10^{45}\rm
ergs^{-1}$, well above the Eddington limit. The X-ray emission
declined as $L_{X}\propto t^{-5/3}$ during the time interval from
$10^{5}$ s to ${10^{6}}$ s after the trigger, while the subsequent
decline can be approximated as $L_{X}\propto t^{-4/3}$
\citep{Levan11a, Levan11b, Bloom11, Cannizzo11}.

The super-long duration ($>$ 8 months) of the X-ray activities
essentially rules out the Gamma-ray Burst models \citep{Shao11}.
Instead it strongly favors the model of tidal disruption of a
(giant) star by a massive black hole \citep{Bloom11, Burrows11,
Levan11b, Cannizzo11, Shao11, Wang12}. As for the central black hole
(BH), it is impossible to measure the mass directly. Indirect
constraints indicate that the mass should be in the range
$10^{6}M_{\odot}-10^{7}M_{\odot}$ (detailed analysis can be seen in
Cannizzo et al. 2011). In the tidal disruption scenario, if a star
passes within the disruption radius $R_{T}\approx
R_{\ast}(M_{BH}/M_{\ast})^{1/3}$, the BH's tidal gravity exceeds the
star's self-gravity and consequently the star is disrupted, where
$R_{\ast}$ ($M_{\ast}$) is the radius (mass) of the disrupted star
and $M_{BH}$ is the mass of the BH. After about hours to weeks, part
of the remnants remains on bound and will return to the pericenter
of the orbit where the material start to be accreted inward,
releasing a flare of energy \citep{Rees88, Rees90, Phinney89,
Strubbe09}.

Theoretical calculations of the tidal disruption events suggest that
the immediately accreted unbinding gas falls back to the pericenter
at a rate $\dot{M}_{\rm fb}\propto t^{-5/3}$ \citep{Rees88,
Phinney89}, and the subsequent disk accretion follows a rate
$\dot{M}_{\rm fb}\tilde{\propto} t^{-4/3}$ if the disk is thick
\citep{Cannizzo09}. So for $t>10^{5}$ s, the X-ray emission
luminosity of Sw J1644+57 is proportional to the accretion rate,
i.e., $L_{X}\propto\dot{M}_{\rm fb}$. In this work we pay special
attention on the physical implication of such a relation. The
physical parameters of the disrupted star are also investigated.

\section{The physical parameters of the disrupted star}

There is an express, derived in the Newtonian limit, for
the timescale of return of the most bound stellar material to the
pericenter \citep{Rees88, Phinney89}
\begin{equation}
t_{\rm fb}=2\pi GM_{BH}(2\Delta E)^{-3/2}
=0.048 ~{\rm yr}(\frac{M_{BH}}{5\times 10^6M_{\odot}})^{1/2}(\frac{M_{\ast}}{M_{\odot}})^{-1}(\frac{R_{\ast}}{R_{\odot}})^{3/2}\mu^3
\end{equation}
where $\Delta
E=k\frac{GM_{\ast}}{R_{\ast}}(\frac{M_{BH}}{M_{\ast}})^{1/3}$, the
dimensionless coefficient $k$ depends on the spin-up state of the
star. If the star is spun up to the break-up spin angular velocity,
we have $k\approx3$. If the spin-up effect is negligible then we
have $k\approx1$ \citep{Rees88, Ayal00}. The dimensionless
coefficient $\mu\equiv R_{p}/R_{T}$ is taken to be a free parameter
in the following discussion, where $R_{p}$ is the periastron
distance of the star.

The star could be spun-up via tidal interaction.
In linear perturbation theory, the spin-up
angular velocity is given by\citep{Press77,Alexander01a,
Alexander01b, Li02}
\begin{equation}
\frac{\omega_{s}}{\omega_{p}}\approx\frac{T_{2}(\mu^{3/2})}{2I\mu^{3}}
\end{equation}
where $\omega_{p}\equiv v_{p}/r_{p}$ is the orbit angular velocity
of the star at the pericenter, $I$ is the stellar momentum of
inertia in units of $M_{\ast}R_{\ast}^2$, $T_{2}$ is the second
tidal coupling coefficient and depends on the structure of the star
and the eccentricity of the orbit.
For an $n=1.5$ polytrope star of mass 0.76 $M_{\odot}$ and radius
0.75 $R_{\odot}$, Alexander $\&$ Kumar(2001) found $I\approx0.21$
and $T_{2}(1)\approx0.36$, corresponding to
$\omega_{s}/\omega_{p}\approx0.86$ for $\mu=1$. Furthermore, they
showed that the numerical simulations including nonlinear effects
led to a larger energy transfer from the orbit to the star and a
larger spin-up than that predicted by linear theory. Therefore we
take $k=3$ as the fiducial value in our analysis. For completeness
we also present the results in the non-spinning case (i.e., $k=1$).

The bound material returns to pericenter at the rate
\citep{Phinney89}
\begin{equation}
\dot{M}_{\rm fb}=\frac{2\Delta M}{3t_{\rm fb}}(\frac{t-t_{s}}{t_{\rm fb}})^{-5/3}
=5.6\times10^{23}~{\rm g~s^{-1}}(\frac{f}{0.1})(\frac{M_{BH}}{5\times10^6M_{\odot}})^{1/3}
(\frac{M_{\ast}}{M_{\odot}})^{1/3}(\frac{R_{\ast}}{R_{\odot}})(\frac{t-t_{s}}{1yr})^{-5/3}\mu^2,
\end{equation}
where $t_{s}$ is the time of initial tidal disruption, $\Delta M$ is
the mass that falls back to pericenter and the dimensionless factor
$f$ is defined as $f\equiv\Delta M/M_{\ast}$.


When the accretion rate of the fall-back material is highly
super-Eddington, only a fraction $(1-f_{out})$ of such material
forms a disc and can be accreted all the way down to the central BH,
i.e., $\dot M=(1-f_{out})\dot M_{fb}$. The remaining part will
instead leave the system undergoing a strong radiation pressure.
\citet{Strubbe09} took a constant $f_{out}$=0.1. However, numerical
simulation indicates that the parameter $f_{out}$ is a growing
function of $\dot{M}_{fb}/\dot{M}_{Edd}$, reaching
$f_{out}\approx0.7$ for $\dot{M}_{fb}/\dot{M}_{Edd}=20$
\citep{Dotan10}. Considering the observed X-ray luminosity
$L_{X}\sim10^{47}-4\times 10^{48}$ erg $\rm s^{-1}$ during the first
$10^6$ s after the trigger (Burrow et al. 2011), even for a
radiation efficiency as high as $0.1$, we need an accretion rate
$\dot M=10L_{X}/c^2\sim5\times 10^{-7}-2\times 10^{-5}\dot
M_{\odot}$. The Eddington luminosity can be scaled as
$L_{Edd}\simeq6.25\times10^{44}M_{6.7}{\rm ~erg ~s^{-1}}$, so the
Eddington rate $\dot{M}_{Edd}\equiv10
L_{Edd}/c^2=3.5\times10^{-7}M_{\odot}~{\rm s^{-1}}$. We then have
$\dot M/\dot M_{Edd}\sim 1-60$.
We define a free dimensionless parameter $\xi=f(1-f_{out})$, the
fraction of the material that is actually accreted onto the central
BH.

Assuming that the jet radiation efficiency is $\epsilon$ during the
stellar debris fallback accretion, the intrinsic jet luminosity then
can be given by
\begin{equation}
L_{\rm j}=\epsilon \dot{M}c^2\approx 2.5\times10^{42} ~{\rm erg ~s^{-1}}
 (\frac{\epsilon}{0.01})(\frac{\xi}{0.05})(\frac{M_{BH}}{5\times10^6M_{\odot}})^{1/3}
(\frac{M_{\ast}}{M_{\odot}})^{1/3}
(\frac{R_{\ast}}{R_{\odot}})(\frac{t-t_{s}}{1yr})^{-5/3}\mu^2.
\end{equation}
where the efficiency is normalized to $\epsilon\sim0.01$ (In sec 3.3
we will show the jet should be launched mainly via magnetic
activities (e.g. B-Z effect) and the efficiency is about 0.01). As
shown below, the conclusion drawn in this section is independent of
the value of $\epsilon$.

When most of bound debris falls back to the pericenter, the jet
luminosity peaks at $t-t_{\rm s}=t_{\rm fb}$ and can be estimated as
\begin{equation}
L_{\rm j,peak}\approx 3.9\times10^{44}~{\rm erg~s^{-1}}(\frac{\epsilon}{0.01})(\frac{\xi}{0.05})(\frac{M_{BH}}{5\times10^6M_{\odot}})^{-1/2}
(\frac{M_{\ast}}{M_{\odot}})^{2}(\frac{R_{\ast}}{R_{\odot}})^{-3/2}\mu^{-3}.
\end{equation}
The observed maximal X-ray luminosity is $\sim 4\times10^{48}$ erg
$\rm s^{-1}$. For a collimated emitting region with a half-opening
angle $\theta_{\rm j}$, we have the constraint $L_{\rm
X,peak}\geq2\times10^{46}(\frac{\theta_{\rm j}}{0.1})^2$ erg $\rm
s^{-1}$. Assuming that most of the radiated energy is in the X-ray
band during the time interval, we have
\begin{equation}
(\frac{\epsilon}{0.01})(\frac{\theta_{j}}{0.1})^{-2}(\frac{\xi}{0.05})(\frac{M_{BH}}{5\times10^6M_{\odot}})^{-1/2}
(\frac{M_{\ast}}{M_{\odot}})^{2}(\frac{R_{\ast}}{R_{\odot}})^{-3/2}\mu^{-3}\geq51.
\end{equation}

The observed X-ray fluence $S_{X}$ ($0~{\rm s}<t<10^{7}~{\rm s}$)
suggests a total energy $\Delta E_X =\theta_{\rm j}^2 \int_{0}^t L_X
(t) dt/2 \sim 1 \times 10^{51}\rm (\frac{\theta_{\rm j}}{0.1})^2
erg$. The total mass of the accreted material is thus
\begin{equation}
M_{\ast}=\frac{\Delta E}{\xi\epsilon c^2}\approx\frac{\Delta
E_{X}}{\xi\epsilon c^2 }\approx1.1M_{\odot}(\frac{\epsilon}{0.01})^{-1}(\frac{\xi}{0.05})^{-1}(\frac{\theta_{j}}{0.1})^2.
\end{equation}
The approximated mass-radius relationship can be scaled as
$(\frac{R_{\ast}}{R_{\odot}})=(\frac{M_{\ast}}{M_{\odot}})^{\eta}$.
For the main sequence stars, we have $\eta\approx0.8$ for
$0.1M_{\odot}<M_{\ast}<1M_{\odot}$  and $\eta\approx0.6$ for
$1M_{\odot}<M_{\ast}<10M_{\odot}$\citep{Kippenhahn94}. With
equations (6) and (7) we obtain
\begin{equation}
(\frac{M_{\ast}}{M_{\odot}})^{1-3\eta/2}\mu^{-3}(\frac{M_{BH}}{5\times10^6M_{\odot}})^{-1/2}\geq46.4,
\end{equation}
which then yields
\begin{equation}
\mu\leq0.36(\frac{M_{BH}}{5\times10^6M_{\odot}})^{-1/6}(\frac{M_{\ast}}{M_{\odot}})^{1/3-\eta/2}.
\end{equation}
Interestingly, the parameter $\mu$ is independent of $\epsilon$,
$\xi$ and $\theta_{j}$. Its dependence on both the stellar mass and
the black hole mass is also very weak. The above analysis is under
the condition that the star is spun-up to the break-up spin angular
velocity, i.e. $k=3$. For the non-spinning case $k=1$, we can obtain
a more stringent result
$\mu\leq0.16(\frac{M_{BH}}{5\times10^6M_{\odot}})^{-1/6}(\frac{M_{\ast}}{M_{\odot}})^{1/3-\eta/2}
$. Therefore, we conclude that the periastron distance is likely
well within the tidal disruption radius (i.e., it is a plunging
event), in agreement with \citet{Cannizzo11}.

Based on the observation data, the peak accretion rate can be
derived with Eq.5 and Eq.7,
\begin{equation}
\dot{M}_{\rm peak}=\frac{\rm L_{j,peak}}{\epsilon c^2}=\frac{\xi M_{\ast}L_{\rm X,peak}}{\Delta E_{X}}=
4.4\times10^{-5}M_{\odot}~s^{-1}(\frac{\xi}{0.05})(\frac{M_{\ast}}{1.1M_{\odot}})
\end{equation}
The accretion rate $\dot{M}$ can be scaled as $\dot{M}=\dot{M}_{\rm
peak}(\frac{t-t_{s}}{1yr})^{-5/3}$ during the fall-back accretion
process.

In this plunging event, the disrupted star's orbit is likely to
mis-align with the equatorial plane of the spinning central BH. A
tilted accretion disk should be formed and the jet aligned with the
disk normal vecter is expected to precess \citep{Stone12, Lei12}.
Saxton et al.(2012) have analyzed the X-ray timing and spectral
variability of Sw J1644+57 and found the periodic modulation,
possibly due to the jet precession.

\section{The radiation mechanism in the fallback phase}
After the time $t\sim10^5$s, the observed X-ray luminosity followed
the fall-back accretion rate (Levan 2011a,b;
Bloom et al. 2011), i.e. $L\propto\dot{M}$. Such a relationship have
shed some light on the underlying physics.

\subsection{Thermal X-ray radiation from the disk?}
While the fallback accretion rate is super-Eddington, the stellar
material returning to pericenter is so dense that it can not radiate
and cool. The gas is most likely to form an advective dominated
accretion flow (ADAF) accompanied with powerful outflow, which
dominates the emission. Most of the radiation will be emitted from
the outflow's photosphere. When the photosphere lies inside the
outflow, the photosphere's radius and temperature can be written as
\citep{Strubbe09, Rossi09}
\begin{equation}
R_{ph}\sim 4f_{out}f_{v}^{-1}(\frac{\dot{M}_{fb}}{\dot{M}_{Edd}})R_{p,3R_{s}}^{1/2}R_{s}
\end{equation}
and
\begin{equation}
T_{ph}\sim 1\times10^{5} ~{\rm K}~ f_{out}^{-1/3}f_{v}^{1/3}(\frac{\dot{M_{fb}}}{\dot{M_{Edd}}})^{-5/12}
M_{6.7}^{-1/4}R_{p,3R_{s}}^{-7/24}
\end{equation}
where $R_{s}\equiv\frac{2GM_{BH}}{c^{2}}$ is the Schwarzschild
radius, and $f_{v}$ is the ratio of terminal velocity of gas with
the escape velocity at the radius $\sim 2 R_{p}$. However, the
photons escape from the photosphere are mainly in the UV optical
band and have a blackbody spectrum. The UV optical emission
luminosity can be scaled as $\nu L_{\nu}\sim 4\pi R_{ph}^2\nu
B_{\nu}(T_{ph})\propto R_{ph}^2 T_{ph} \propto
\dot{M_{fb}}^{19/12}$.

These optical photons could be Compton-scattered by the relativistic
electrons in the outflow. The energy of the photons getting
scattered is $h\nu_{IC}\approx D^{2}\gamma^{2}h\nu$, where
$D=1/[\Gamma(1-\beta cos\theta)]$ is the Doppler factor, $\Gamma$ is
the Lorentz factor of the outflow, $\theta$ is the angle between the
outflow axis and the observer's line of sight, $\gamma$ is the
Lorentz factor of the relativistic electrons. The energy of inverse
Compton scattered photons can peak in the X-ray band if the
parameter $D\sim1$ and $\gamma\sim 10$. However even in this case,
the X-ray luminosity does not satisfy the relation $L_{x}\propto \dot{M}_{fb}$,
inconsistent with the observational data.

\subsection{Neutrinos annihilation launched outflow?}
When the mass accretion rate is high enough, the accretion proceeds
via neutrino cooling and neutrinos can carry away a significant
amount of energy from the inner regions of the disk. The mechanism
is used to explain the launch of at least some gamma-ray burst
outflows \citep{Popham99,Fan11,Fan12}. The luminosity may be well
approximated by a simple formula \citep{Zalamea11}

\begin{eqnarray}
      L_{\nu\bar{\nu}} & \approx & 1.1\times 10^{52}\,\chi_{\rm ms}^{-4.8}\,
               \left(\frac{M_{BH}}{3M_\odot}\right)^{-3/2} \\
      & \times  & \left\{ \begin{array}{ll}
                    0                  &  \dot{M}<\dot{M}_{ign} \\
                \dot{m}^{9/4} \;\;     &  \dot{M}_{ign}<\dot{M}<\dot{M}_{trap} \\
          \dot{m}_{\rm trap}^{9/4}\;\; &  \dot{M}>\dot{M}_{trap} \\
                           \end{array}
                  \right\}             \,{\rm erg~s}^{-1},
\end{eqnarray}
where $\dot{m}=\dot{M}/M_{\odot}$~s$^{-1}$, $\chi_{\rm
ms}=R_{ms}(a)/R_{s}$, $R_{ms}$ is the radius of the marginally
stable orbit, $\dot{M}_{ign}$ is the mass accretion rate to ignite
the neutrino emitting, $\dot{M}_{trap}$ is the mass accretion rate
when the emitted neutrino becomes trapped in the disk and advected
into the black hole. The characteristic accretion rates
$\dot{M}_{ign}$ and $\dot{M}_{trap}$ depend on the viscosity
parameter $\alpha$ and the mass of central BH. 
Based on the work of \citet{Beloborodov03}, for $\nu$-transparent,
the accretion rate should be as large as
$\dot{M_{ign}}>7.6\times10^{30}(\frac{r}{3r_{s}})^{1/2}(\frac{\alpha}{0.1})
(\frac{M_{BH}}{M_{\odot}})^2{\rm g~s^{-1}}$. For this event Sw
J1644+57, a plunging one, the peak accretion rate is $\dot{M}_{\rm
peak}\approx4.4\times 10^{-5}M_{\odot}~{\rm s^{-1}}$, which is far
less than $\dot{M}_{ign}$. Hence we conclude that the observed X-ray
emission could not be produced by neutrino annihilation (please see
Shao et al. (2011) for an alternative argument disfavoring the
neutrino mechanism).

\subsection{Poynting-flux dominated outflow?}
Extracting energy from the rotating black hole may be possible
through the Blandford-Znajek mechanism \citep{Blandford77}. Such a
process is based on the expectation that the differential rotation
of the disk will amplify pre-existing magnetic fields until they
approach equipartition with the gas kinetic energy. For a black hole
of mass $M_{BH}$ and angular momentum $J$, with a magnetic field
$B_{\bot}$ normal to the horizon at $R_{h}$, the power arising from
BZ mechanism is given by (e.g. Thorne et al. 1986)
\begin{equation}
L_{BZ}=\frac{\pi}{8}\omega_{F}^2(\frac{B_{\bot}^{2}}{4\pi})R_{h}^2c(\frac{J}{J_{max}})^2,
\end{equation}
where $J_{max}=GM^2/c$ is the maximal angular momentum of the black
hole. The factor
$\omega_{F}^2=\Omega_{F}(\Omega_{h}-\Omega_{F})/\Omega_{h}^2$
depends on the angular velocity of field lines $\Omega_{F}$ relative
to that of the black hole, $\Omega_{h}$. Usually we adopt
$\omega_{F}=1/2$, which maximizes the power
output\citep{Macdonald82, Thorne86}. We follow the common assumption
that the magnetic field in the disk will rise to some fraction of
its equipartition value $P_{mag}=\frac{B^2}{8\pi}\sim \alpha P$ in
the inner disk. The pressure $P=\rho c_{s}^2$ is given by
\citep{Armitage99}
\begin{equation}
P=\frac{\sqrt{2}\dot{M}}{12\pi\alpha}(5+2\varepsilon)^{1/2}(GM)^{1/2}R^{-5/2},
\end{equation}
where $\varepsilon$ is the parameter governing the property of the disk. For a thick disk we have
$\varepsilon<1$ otherwise $\varepsilon>1$. In the inner region of the disk, we
assume $B_{\bot}\approx B$, $R\approx R_{h}=GM/c^2$. The
BZ power for the case of a maximally rotating black hole (i.e.,
$J=J_{max}$) can be estimated as
\begin{equation}
L_{BZ}\approx 7\times10^{-3}(5+2\varepsilon)^{1/2}\dot{M}c^2,
\end{equation}
corresponding to an efficiency
$\epsilon_{BZ}=L_{BZ}/\dot{M}c^2\sim10^{-2}$ for the thick disk
model. In the thin-disk scenario the radiative efficiency can be
as high as $\sim 0.1$. For Sw J1644+57, at the time
$t-t_{s}=t_{fb}+10^6$s, the mass accretion rate
is about $3.3\times 10^{-8}M_{\odot}~{\rm s^{-1}}$, the observed luminosity
is $2L_{BZ}\theta_{j}^{-2}\sim5\times10^{47}\rm erg$ ${\rm
s^{-1}}(\theta_{j}/0.1)^{-2}$, consistent with the observation.

Our conclusion that the outflow powering the super-long X-ray
emission should be launched via magnetic activities (e.g., B-Z
mechanism) is consistent with that of Shao et al. (2011). Lei \&
Zhang(2011) have also analyzed the jet launched by B-Z mechanism and
then constrained the physical parameter of the central BH. One
interesting finding is that the central BH should have a moderate to
high spin.

In the Poynting-flux dominated outflow, the X-ray emission could be
due to the dissipation of the magnetic field \citep{Usov94,
Thompson94}. There are several magnetic field dissipation models
that could produce the observed emission, such as the global MHD
condition breakdown model \citep{Usov94}, the gradual magnetic
reconnection model \citep{Drenkhahn02}, the magnetized internal
shock model \citep{Fan2004}, and the collision induced reconnection
model \citep{Zhang11}. For illustration, here we take the global MHD
condition breakdown model to calculate the emission. By comparing
with the pair density ($\propto r^{-2}$, $r$ is the radial distance
from the central source) and the density required for co-rotation
($\propto r^{-1}$ beyond the light cylinder of the compact object),
one can estimate the radius at which the MHD condition breaks down,
which reads \citep{Usov94, Zhang02}
\begin{equation}
r_{MHD}\sim5\times10^{20}~{\rm cm}(\frac{L}{10^{47}})^{1/2} (\frac{\sigma}{10})^{-1}( {\frac{t_{v,m}}{10^2}})
(\frac{\Gamma}{10})^{-1},
\end{equation} where $\sigma$ is the ratio of the magnetic energy
flux to the particle energy flux, $\Gamma$ is the bulk Lorentz
factor of the outflow, $t_{v,m}$ is the minimum variability
timescale of the central engine. Beyond this radius, intense
electromagnetic waves are generated and outflowing particles are
accelerated  (e.g. Usov 1994). Such a significant magnetic
dissipation process converts the electromagnetic energy into
radiation.

At $r_{MHD}$, the corresponding synchrotron radiation frequency can
be estimated as \citep{Fan05, Gao06}

\begin{equation}
\nu_{m,MHD}\sim1.5\times10^{18}~{\rm Hz}~(\frac{1+z}{1.35})^{-1}(\frac{\zeta}{0.1})
C_{p}^{2}(\frac{\sigma}{10})^3(\frac{\Gamma}{10})(\frac{t_{v,m}}{10^2}),
\end{equation}
where
$C_{p}\equiv(\frac{\epsilon_{e}}{0.1})[\frac{13(p-2)}{3(p-1)}]$,
$\epsilon_{e}$ is the fraction of the dissipated comoving magnetic
field energy converted to to the comoving kinetic energy of the
electrons, and the accelerated electrons distribute as a single
power-law $dn/d\gamma_{e}\propto\gamma_{e}^{-p}$, $\zeta<1$ reflects
the efficiency of magnetic energy dissipation. So most energy is
radiated in the X-ray band.



\section{Clue to the X-ray steep decline following the prompt emission in gamma-ray bursts}
The observations of Sw 1644+57 suggest that the long-lasting
fall-back accretion onto a black hole can produce energetic X-ray
emission and the radiation luminosity traces the accretion rate
(i.e., $L_{X}\propto \dot{M}$). One interesting question is whether
similar process takes place in Gamma-ray Bursts (GRBs) or not. The
answer may be positive. Here we just discuss the X-ray steep decline
(quicker than $t^{-3}$, see Fig.1 of Zhang et al. (2006) for
illustration) following the prompt emission in GRBs. In the
collapsar model, a fraction of the gas in the core of the collapsing
star has not sufficient centrifugal support and directly forms a
central black hole. The rest of the material will have sufficient
angular momentum to go into orbit around the black hole. The
fall-back accretion rate is tightly related with the pre-collapse
stellar density profile, which is of the form $\rho\propto
r^{-\tau}$. Numerical simulation has found when the outermost 0.5
$M_{\odot}$ layer of the star (where $\tau >5$) is accreted, the
fall-back accretion rate can get to the value $\dot{M_{fb}}\propto
t^{-3}$ or steeper \citep{MacFadyen01, Kumar08}. If the relation
$L_{X}\propto \dot{M}$ still holds, one has an X-ray emission
decline steeper than $t^{-3}$, in agreement with the observational
data  \footnote{\bf An alternative scenario is that
the central engine turns off abruptly and the quick decline is dominated
by the high latitude emission of the prompt emission pulses
(see Zhang et al. 2006 and the references therein). With the future X-ray
polarimetry data we may be able to distinguish between these
two kinds of models if our line of sight is not along the center of the ejecta (Fan et al. 2008).}.

\section{Conclusion and Discussion}
Swift J164449+573451 is a peculiar outburst which is most-likely
powered by the tidal disruption of a star by a massive black hole.
In this work we find out that the ratio of the periastron distance
to the disruption radius $\mu<1$, implying that SW J1644+77 is a
plunging event (see section 2). The mass of the plunging star
however can not be tightly constrained due to its strong dependence
on the poorly understood parameters $\epsilon$ (the radiation
efficiency), $\xi$ (the fraction mass of the star that is eventually
accreted into the central) and $\theta_{j}$ (the half opening angle
of the collimated outflow).

As a tidal disruption event, the accretion rate $\dot{M}$ at late
times (say, $t>10$ day) is relatively well understood and is widely
believed to be $\propto t^{-4/3}$. The detected X-ray emission
$L_{X}$ shows a rather similar decline behavior. Since the forward
shock origin of the long-lasting and highly variable X-ray emission
has already been convincingly ruled out (Shao et al. 2011), the
X-ray emission has to be from an outflow launched by the accreting
black hole. These two facts strongly suggest that
$L_{X}\propto\dot{M}$, which can shed valuable light on the
underlying physics, in particular the energy extraction process.
Three kinds of possible mechanisms have been examined and only the
Poynting-flux dominated outflow model is found to be able to account
for the data (see section 3 for details). Therefore the magnetic
activity at the central engine (e.g., Blandford-Znajek process or
Blandford-Payne process) plays the main role in extracting the
rotation energy of the black hole and then launching the outflow. We
suggest that $L_{X}\propto\dot{M}$ may also hold in the quick
decline phase of Gamma-ray Bursts.

\section*{Acknowledgments}
We thank the anonymous referee for constructive comments and Dr.
Yizhong Fan for his kind help in improving the presentation. This
work was supported in part by the National Natural Science Foundation of
China under the grant 11073057 and by the Key Laboratory of Dark
Matter and Space Astronomy of Chinese Academy of Sciences.

\clearpage

\end{document}